\documentclass[11pt,twoside]{article}
\usepackage{./asp2014}

\aspSuppressVolSlug
\resetcounters

\begin{document}

\title{Some Remarks About Solar System Structure, Models, Observations and Wing Ip}
\author{David Jewitt$^1$}
\affil{$^1$University of California at Los Angeles, Los Angeles, CA, USA; \email{jewitt@ucla.edu}}

\paperauthor{David Jewitt}{Author1Email@email.edu}{}{UCLA}{Earth, Planetary and Space Sciences}{Los Angeles}{CA}{90095}{USA}
 
\begin{abstract}
This is an opinion-piece based on a talk given at the Summer 2017 Serendipities in the Solar System (``Ip-Fest'') meeting in Taiwan\footnote[2]{A celebration of Wing Ip's 70th birthday, held at the Institute of Astronomy, National Central University, Taiwan, July 2017; published in 2018 as Astronomical Society of the Pacific Conference Series Volume No.~513, page 33}. I begin with an overview of the new solar system, then discuss modelling attempts, then  the  distribution of optical colors  as a proxy for the distribution of compositions, and I end with remarks about Wing Ip. 
\end{abstract}

\section{Introduction}
Three cometary reservoirs exist (see  Figure \ref{schematic} and Jewitt et al.~2009 for a review):

1) The smallest is the asteroid-belt itself which, despite its high radiation equilibrium temperatures,  preserves bulk ice (see also Snodgrass et al.~2017).  In the main-belt comets, ice is  shielded from the Sun by refractory mantles and only occasionally sublimates when exposed by impact or other surface disturbances. The duty cycle (ratio of ``on'' time to total elapsed time) is $< 10^{-4}$.  There are about 10$^6$ asteroids larger than a kilometer in scale.  The fraction containing ice is uncertain, but may approach unity in the outer-belt.  

2) The largest reservoir is the Oort cloud, a spherical swarm of vast extent ($\sim10^4$ to $10^5$ AU scale) from which the long-period comets are supplied (see also Rickman 2014).  The number of Oort cloud comets is uncertain, but is probably in the 10$^{11}$ to 10$^{12}$ range.  Their combined mass  is likely in the 1$M_{\oplus}$ to 10$M_{\oplus}$ range (1$M_{\oplus}$ = 6$\times$10$^{24}$ kg), but could be a little smaller or much larger, depending mainly on the unmeasured properties of the  inner Oort cloud.  

3) The third cometary reservoir is the Kuiper belt (a.k.a.~the trans-Neptunian belt), source of most short-period comets.  The Kuiper belt holds at least 10$^9$ objects larger than 1 km and 10$^5$ objects larger than 100 km, with a combined mass not more than $\sim$0.1$M_{\oplus}$.  

The three reservoirs sample  products of low temperature accretion in different radial regions of the protoplanetary disk.  The main-belt comets likely formed in-place, close to the snow-line in the epoch of accretion, although other formation locations have been suggested.  The Oort cloud comets probably formed in the middle solar system, where we now find the giant planets, and were subsequently hurled outwards by near-miss gravitational scattering from the planets as they grew.  Most were lost to the interstellar medium, never to be seen again, but maybe 1\% to 10\%~of the comets were deflected on the way out  by the combined gravitational effects of passing stars and the galactic tide. They became trapped in long-lived, weakly-bound heliocentric orbits with perihelia far outside the planetary region (e.g.~Dones et al.~2015).  The Kuiper belt contains a mixture of objects that we think were scattered into place from interior source regions (the so-called dynamically hot populations) with objects that likely formed in-place (the cold classical objects).  The reservoirs thus offer a kind of radial stratigraphy of the solar system, by providing icy products accreted at temperatures $\sim$140 K (main-belt comets) to $\sim$40 K (cold classicals).  

\articlefigure{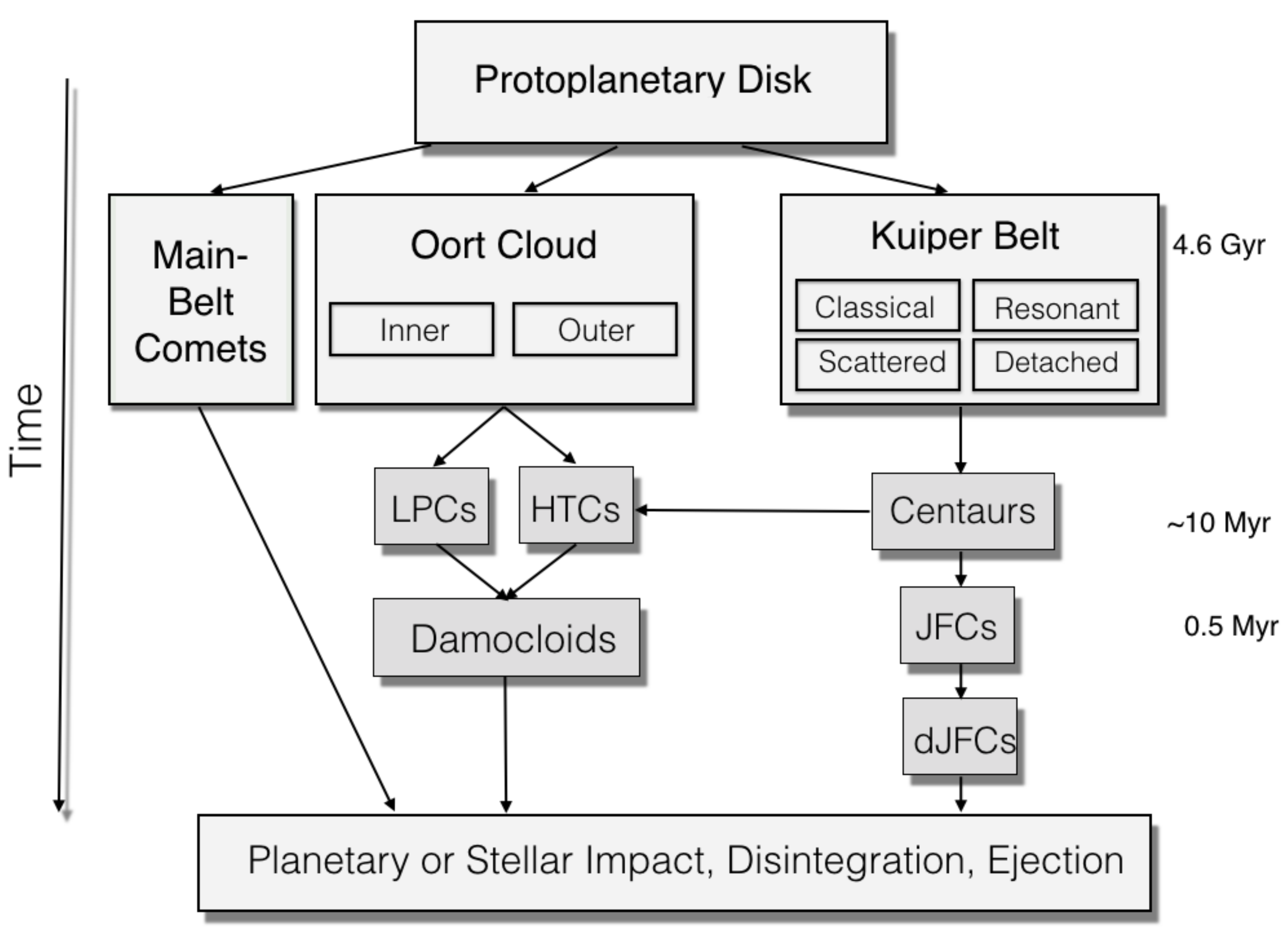}{schematic}{Schematic diagram showing the relationships believed to exist between various small-body solar system populations. Acronymns in the Figure indicate different sub-populations: LPCs = Long-period comets, HTCs = Halley-type comets, JFCs = Jupiter family comets, dJFCs = dead (or dormant) Jupiter family comets.  Damocloids are probably dead (or dormant) HTCs.  Numbers on the right indicate that, while the reservoirs survive for the age of the solar system,  bodies outside the reservoirs meet a rapid demise.  Bodies scattered by the giant planets have median lifetimes $\sim$10$^7$ yr while those in the cramped quarters of the terrestrial planet region have dynamical lifetimes $\sim$0.5 Myr (and physical lifetimes considerably shorter).  All comets displaced from their reservoirs meet similar fates.}

Study of the Oort cloud is limited by its vast size, which renders its constituent comets  too faint  to be detected from Earth. We are forced to infer the properties (and existence!) of the cloud from comets that have left it (the problem is analogous to trying to figure out the size and layout of the parking structures at LAX by only counting the number of cars leaving the airport).  Study of the main-belt comets is limited by the small number (only a handful) of known examples (the analogy is trying to figure out the parameters of LAX parking by staring at just one or two parked cars), although there is hope that intensive future observations will change this circumstance.  In contrast, the Kuiper belt is eminently observationally accessible, and the physical and dynamical properties of large numbers of Kuiper belt objects can be measured directly.  Indeed, about 2000 Kuiper belt objects are known at the time of writing and from them the incredible and unexpectedly complicated  architecture of the outer solar system has been revealed.   \textit{The special feature of the Kuiper belt, and the reason that its discovery has revolutionized planetary science, is that it is near-enough for its contents to be studied directly}.

Measurements of the orbits of Kuiper belt objects have provided the food for  modellers interested in the dynamical aspects of the origin and evolution of the solar system.  The key observational finding was that the  abundance of resonant objects (particularly the 3:2 ``Plutino'' population) is too large to be a result of chance.  Instead, the resonant populations are best understood as a consequence of the radial migration of Neptune during the formation epoch (Fernandez and Ip 1984, Malhotra 1995).   If Neptune's orbit changed size, then so must have the orbits of the other planets, with potentially profound dynamical consequences that depend on the degree and the rates (and smoothness) of the migration.  Extreme effects might result if the giant planets were driven to cross mean-motion resonances with each other, as described in the famous Nice model and its variants, derivatives and follow-ons.  Significantly, such effects might include the capture of objects scattered from the Kuiper belt into dynamical niches including the Jovian Trojans (Morbidelli et al.~2005), the irregular satellites (e.g.~Nesvorny et al.~2014) and even the asteroid belt (Levison et al.~2009).  

Most recent solar system models are numerical N-body simulations, relying on fast computers to represent complex dynamical systems.  The biggest strength of the  numerical  models is their flexibility; their many parameters can be readily adjusted to fit many (but not all) of the known properties of the solar system\footnote{An obvious exception is the cold-classical Kuiper belt which, with its low mass, narrow inclination distribution, high abundance of binaries and sharp outer edge, appears in the Nice model and its variants as an ad-hoc addition, not as a natural consequence of planetary migration into a pre-existing planetesimal belt (c.f.~Fraser et al.~2014).}. Their biggest weakness  is exactly the same thing - their flexibility, which is so great that they struggle to offer any firm, observationally testable predictions. The result is scientifically frustrating in the sense that interesting models come and go (the Nice model, the modified Nice model, the jumping Jupiter model, the Grand Tack model, models in which extra giant planets are added to the solar system then allowed to escape, models in which extra components are added to the asteroid belt and then dynamically destroyed) but nobody can tell whether they describe the real world, or just a model-world.   

Consider the  late-heavy bombardment (LHB) of the Moon at 3.8 Gyr as a case in-point.  The  Nice model  was originally presented as a solution to the puzzle of why the LHB was delayed for 800 Myr following the Moon's formation (Gomes et al.~2005).  It did this rather ingeniously, by slowly driving the planets towards a resonant instability, using  torques exerted by a Kuiper belt selected to have the ``right'' mass and the ``right'' separation from the outermost planet.  However,  from the start, the existence of the LHB was doubted by the community best equipped to judge it (geologists and geochemists who studied the Apollo lunar samples, see Hartmann 1975), a fact that was forgotten, ignored or disputed by  the dynamicists.  Improved measurements of lunar rocks (especially the highly refractory and isotopically revealing nuggets called ``zircons'') have strengthened the alternate explanation, namely that the impact flux declined smoothly over a long period of time, not in a spike-like LHB (see Zellner 2017, for a recent review).  In other words, the very problem that the Nice model was proposed to solve has evaporated.  The response of the Nice modelers has been to simply  change a few assumptions about the initial conditions of the solar system in order to push the LHB closer to the formation epoch (Morbidelli et al.~2017).   Having no evidence for the LHB could simply mean, in the context of the model, that resonance crossing occurred so close in time to the formation of the planets that it is indistinguishable from the heavy cratering flux in the accretion phase.  This might be true, of course, but it might also be true that the LHB never occurred and that the Nice model, as constructed, describes something that didn't happen.  The fundamental problem is that the numerical models  can be tuned to provide a wide range of outcomes, but they don't offer the means to test the veracity of those outcomes.

\section{Color Distributions}

The optical reflectivity spectra of most bodies beyond the main asteroid belt are linear functions of wavelength.  We can adequately describe such spectra with a simple color index, B-R, plotted as histograms in Figure (\ref{histo}) (the figure is an improvement on a figure in Luu and Jewitt (1996), which itself has been updated in the literature many times since).  The color of the Sun is B-R = 0.99, so it is immediately obvious from the Figure that essentially everything in the outer solar system is redder than the Sun. 

Objects with B-R $\ge$ 1.6 are described as ``ultrared'' (Jewitt 2002).  The nature of the ultrared material is unclear, but an association with irradiated, complex organics is likely (Dalle-Ore et al.~2015).  The physical picture is of a thin surface layer, roughly 1 m thick (GeV protons penetrate to about this depth), which has been heavily processed by prolonged cosmic ray bombardment leading to the breakage of bonds (hence the featureless spectra), to the escape of hydrogen and to progressive carbonization.  Material beneath this surface skin would be relatively unprocessed and, perhaps, spectrally distinct.

\articlefigure[width=.6\textwidth]{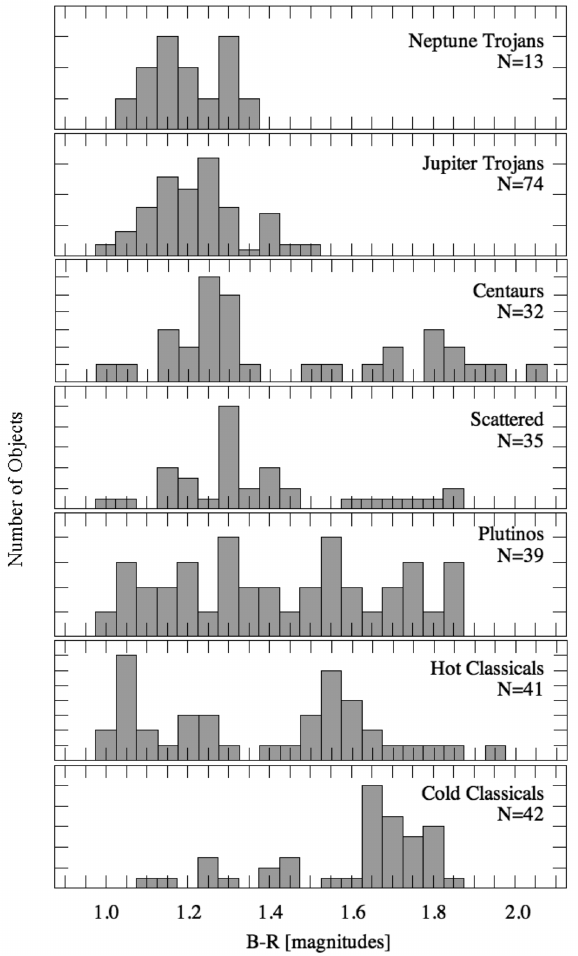}{histo}{Histogram of the B-R color index for outer solar system populations, as labelled.  The number of objects in each sample is listed. From Jewitt 2018. }

\subsection{Kuiper belt}
Kuiper belt objects can be usefully  divided on the basis of their orbits into four distinct types.  The resonant objects, mentioned above, occupy mean-motion resonances with Neptune that provide long-term dynamical stability.  Scattered objects have perihelia close enough to Neptune ($q \lesssim$ 40 AU) that they can be dynamically excited by interactions with that planet.  On solar-system timescales these Neptune interactions generate eccentric orbits that can reach very large distances (for example, the current orbit of 2014 FE72 has aphelion $Q$ = 3800 AU) and may eventually lead to escape from the solar system.  The detached objects are like the scattered objects, but have perihelion distances that are thought to be too large for their orbits to be strongly affected by Neptune in the age of the solar system.  This critical perihelion distance is commonly taken to be $q \ge$ 40 AU but may be larger.  Detached KBOs are sometimes assumed to have been emplaced by perturbations from a passing star, or by other unseen massive perturbers.  The classical objects have semimajor axes $\sim$43 AU and are long-term stable because their eccentricities are modest ($e \lesssim$0.1), preventing close approach to Neptune.  The classical objects are subdivided on the basis of their inclinations, $i$, into cold-classicals ($i <$ 5\deg, although sometimes other critical inclinations are picked) and hot-classicals ($i \ge$ 5\deg).  Here, ``hot'' and ``cold'' refer to the kinetic theory, in which temperature is a measure of the kinetic energy of particles in a distribution.  The hot-classicals and other high mean inclination populations (the scattered objects, the resonant objects) are held to have been dynamically excited relative to the low-inclination cold-classicals (although even these occupy a disk thicker than expected from accretion physics and must have experienced some past disturbing event).
  
It is convenient to compare the color distributions of the different populations in terms of the fraction of each population consisting of ultrared bodies.  
 The histograms of Figure (\ref{histo}) show immediately that the cold classical objects are, on average, the reddest objects with about 2/3rds being ultrared (with a bluer tail extending towards nearly neutral colors).  Ultrared material is present in the other plotted components of the Kuiper belt, including the hot classical objects, the Plutinos and the scattered KBOs, but the fractional abundance in each of these is smaller, closer to 1/3rd. It is also present in the Centaurs (see Section \ref{CandT}). Ultrared matter is not present at all in inner solar system populations. It is absent from the Jovian and Neptunian Trojans and from even the reddest (D-type)  asteroids in the main-belt.   

\subsection{Centaurs and Trojans}
\label{CandT}
Members of the hot population (represented in color space by the above-mentioned hot classicals, resonant objects and scattered disk objects)  escape the Kuiper belt and may come to reside in other regions of the solar system.  The Centaurs are the first stage of these escaped Kuiper belt objects.  The similarity between the color histograms of the Centaurs and the hot populations (Figure \ref{histo}) is compatible with this idea.  Specifically, the Centaurs include an ultrared fraction $\sim$1/3 that is comparable to that seen in the hot populations, as expected if  the Centaurs are escaped objects from the hot population (most likely from the scattered disk).  

\articlefigure[width=.7\textwidth]{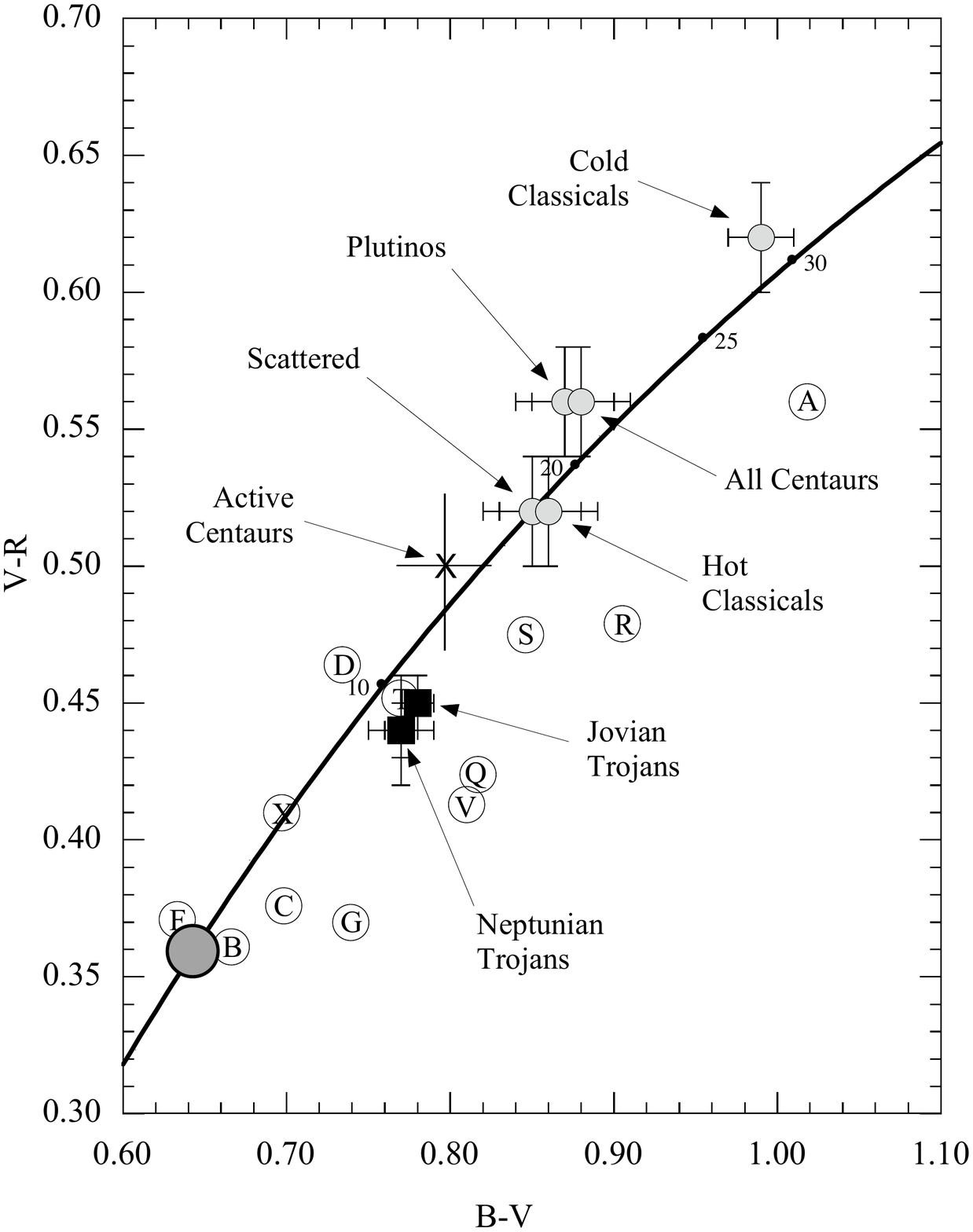}{colorcolor}{Color vs.~color plot for the outer solar system.  Individual outer solar system populations  are labeled.   Grey filled circles show Kuiper belt populations.  Black filled squares show the Jovian and Neptunian Trojans, as marked.  The Centaurs and active Centaurs are plotted, the latter marked X. Circled black letters show the locations of various main-belt asteroid spectral types.   The large shaded circle at lower left marks the color of the Sun.  The black arc shows the locus of points having linear reflectivity gradients (i.e.~being redder than the Sun, by $S'$ [\%/1000\AA]~in wavelength).  Small numbers along the arc show the magnitude of the reflectivity gradient, in \%/1000\AA. Modified from Jewitt (2015).}

On the other hand, the colors of the Trojans of Jupiter are \textit{not} compatible with those of the hot population, counter to expectations if the Trojans are captured KBOs.  In fact, the Jupiter Trojans are completely lacking in ultrared matter.  This observation, known for a long time (Figure 8 of Luu and Jewitt 1996), has been interpreted to mean that the Trojans have experienced surface modification after being displaced from the Kuiper belt, specifically by losing their ultrared matter.

The color data are shown in a different form in Figure (\ref{colorcolor}).  In that Figure, each point represents the mean color (and 1$\sigma$ standard error) on the color of a given population. The black arc shows  the locus of points having linear reflectivity spectra, with gradients increasing from $S'$ = 0\%/1000\AA~at the Sun to $S'$ = 40\%/1000\AA~at the upper right (see the small numbers along the arc).  On the reflectivity gradient scale, $S'$ = 0\%/1000\AA~corresponds to the color of the Sun, B-R = 0.99, while ultrared colors (B-R = 1.6) begin at $S'$ = 25\%/1000\AA.

Figure (\ref{colorcolor}) shows a spectacular dispersion in the mean colors of different populations and it also shows that the reflection spectra are basically linear with wavelength (because they are nicely distributed along the arc).  The linearity reflects the well-known observation that the spectra of most middle- and outer-solar system bodies are comparatively featureless, lacking the strong mineral absorption bands that are common in the rocky asteroids. It also shows that the various KBO groups have the reddest mean colors (with the cold-classicals being the ultrared stand-outs), and that purportedly related populations (e.g.~the Trojans, the Jupiter family comet nuclei) are, for the most part, less red in their mean colors.

A closer look at the Centaurs provides some support for the idea that surface colors (and compositions) are modified upon approach to the Sun.  In Figure (\ref{colorcolor}) I have split the Centaurs into those that are active (showing comet-like comae) and those that are not.  Notice that the active Centaurs are less red than the inactive ones, a difference that is significant at about the 3$\sigma$ level of confidence (Jewitt 2009, 2015).   

\articlefigure[width=0.8\textwidth]{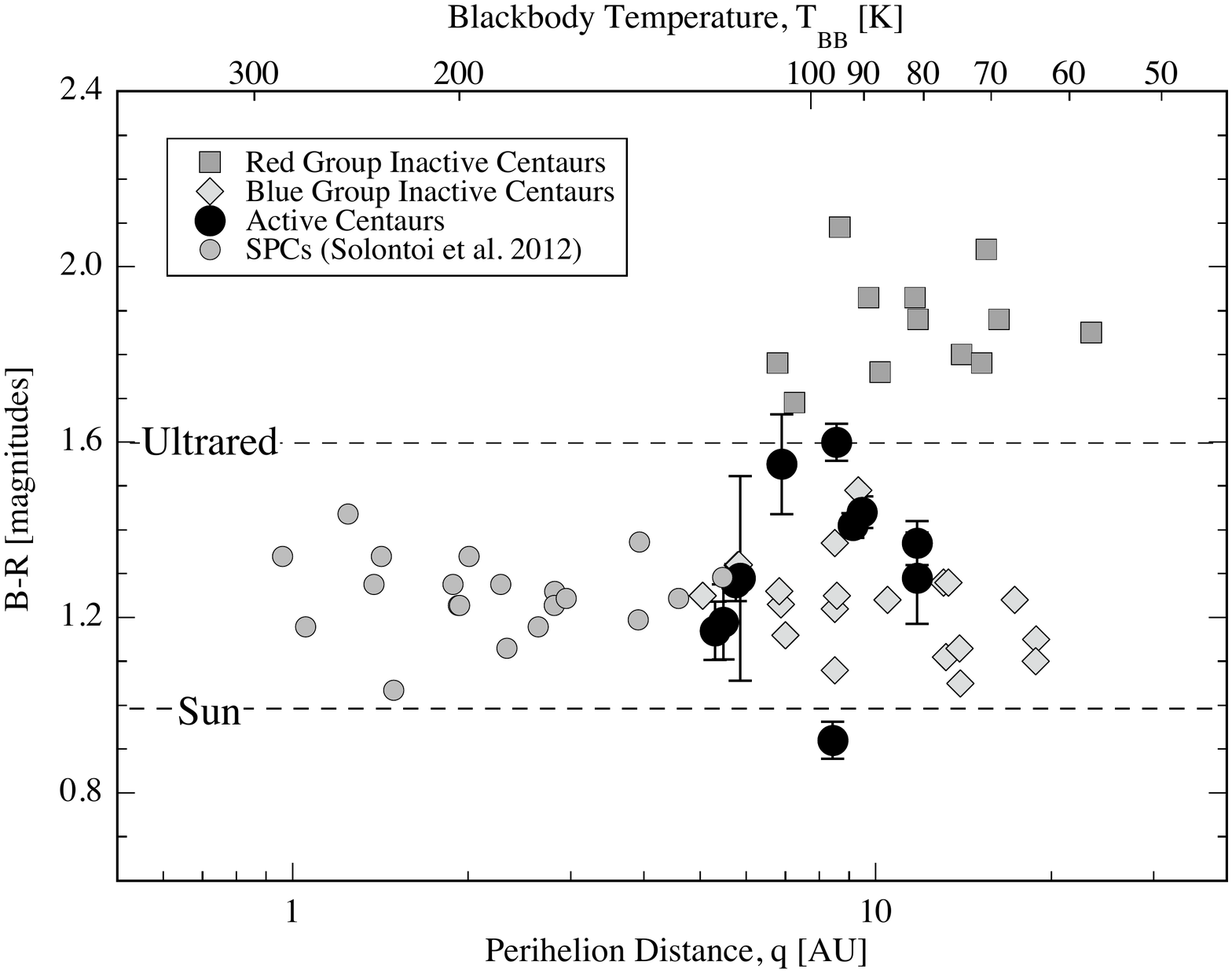}{centaurs}{Centaur colors vs.~perihelion distance. The colors of the Sun and the nominal ultrared matter boundary are marked. Centaur colors are bimodal at  $q >$ 8 AU to 10 AU, with many ultrared (B-R $>$ 1.6) objects, but the ultrared objects are absent at smaller $q$.  From Jewitt 2015.}

The distribution of Centaur colors with perihelion distance, $q$, is even more interesting (Figure \ref{centaurs}).
The Centaur colors are bimodally distributed for $q \gtrsim$ 10 AU (c.f. Peixinho et al.~2012) but unimodal on Centaurs with smaller perihelia and on the Centaurs captured by Jupiter as short-period comets.  It looks like the ultrared matter cannot survive on objects at $q \lesssim$ 10 AU. Since there is no dynamical process that can separate Centaurs by their color, we have to conclude that this is an evolutionary effect.    

What could be the cause of this change in colors inside the orbit of Saturn?  The  blackbody temperature at 10 AU is $\sim$88 K. A clue is provided by the observation that some  Centaurs  become active when $q \lesssim$ 10 AU, the same distance at which the incoming ultrared matter begins to vanish.  Cometary activity on incoming Centaurs could itself be the cause of the loss of ultrared matter by resurfacing, in which some of the ejected material leaves the surface at less than the local gravitational escape velocity and falls back to the surface.  The timescale for building up a ballistically deposited layer is very short, comparable to the orbit period (Jewitt 2002, 2009, 2015) and provides a natural mechanism by which an irradiated organic mantle could be hidden from view by even weak outgassing activity.  Sublimation of widespread surface volatiles could also be responsible, with a particular suggestion being offered that H$_2$S, if present, could sublimate away on objects approaching within $\sim$15 AU of the Sun (i.e.~somewhat farther than the critical distance for the onset of activity and the loss of ultrared surface matter; Wong and Brown 2017).  Whatever the mechanism, the introduction of additional physical processes means that we cannot use the more neutral colors of the Trojans (and comet nuclei, for that matter) as evidence against the suggestion that these bodies originated in the Kuiper belt hot population.  The situation is more interesting for the Neptunian Trojans, whose B-R distribution is indistinguishable from that of the Jovian Trojans (Figure \ref{histo}, and Jewitt 2018).  At 30 AU (and 50 K), the Neptune Trojans arte far too cold for Centaur-like activity to occur. Indeed, their radiation equilibrium temperatures are negligibly different from those of the Kuiper belt objects beyond.

\subsection{Comets}
\articlefigure[width=0.8\textwidth]{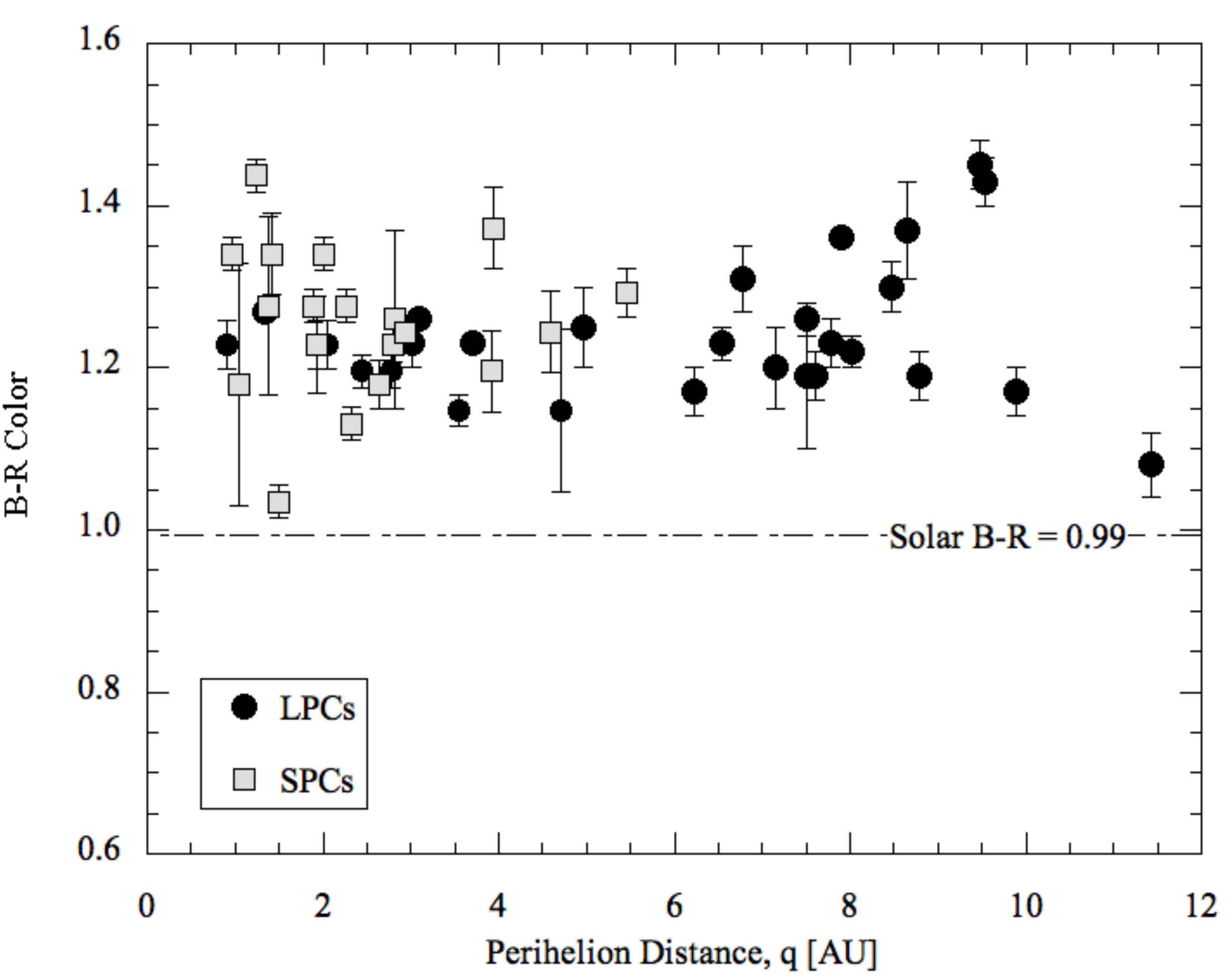}{comets}{Colors of comets (long-period and short-period, as marked) vs.~perihelion distance. The comets are all redder than the Sun but none is ultrared (B-R $>$ 1.6).  Long-period and short-period comets have identical color distributions.  From Jewitt 2015.}

The optical colors of comets are remarkably concordant (Figure \ref{comets}).  Long-period (Oort cloud) and short-period (Kuiper-belt) comets are indistinguishable, color-wise.  No comets have been found to contain ultrared matter.  It is sometimes asserted that the colors of active comets are affected by geometrical scattering effects in particles with dimensions comparable to the wavelength of light (the appropriate dimensionless parameter is $x = 2\pi a/\lambda$, where $a$ is the particle radius and $\lambda$ is the wavelength).  Tiny particles with $x \ll 1$, for example, tend to scatter blue light more efficiently than red.  Large particles ($x > 1$) scatter with an efficiency that is more reflective (pun intended) of their composition than of their size or shape.  However, the cometary dust size distributions are such that the mean size for scattering is always  $x > 1$,  therefore non-geometric scattering effects are not dominant and there are no blue comets.  The absence of ultrared matter, even on dynamically new comets arriving fresh from the Oort cloud where irradiated organics might be expected, could mean that the ejected dust emanates from layers beneath the irradiation layer.  The uniformity of the colors between long- and short-period comets  is compatible with the  scattering of dust across a wide range of heliocentric distances in the protoplanetary disk, as evidenced separately by the incorporation of crystalline silicates in comets (Ogliore et al.~2009).  It is also compatible with a large degree of overlap between the radial distances of formation of the LPCs and SPCs.

\section{Wing Ip}
This informal note (together with the Ip-fest talk on which it is based) celebrates  Wing Ip's constructive influence on planetary science over many years. It would be remiss of me not to include at least a few remarks about Wing as a person.  

First, an anecdote; I don't know what it means, but it happened.  Wing gave a talk at the University of Hawaii while I was employed there, perhaps 15 years ago.  I went to his talk, which was about the Kuiper belt.  While I forget most of the details  I do remember that about 20 minutes into his presentation, Wing stopped and stared at me with a strange look in his eye.  "Would you like to say something?", he asked.  "No", I replied.  "Why don't you talk for a while?", he continued.  "Er...because it's your talk, Wing, not mine", I replied.  "Just 10 or 15 minutes, just a little bit", he urged in his usual, genial way.  So I stood up and talked about the Kuiper belt for 10 or 15 minutes while he sat down and appeared to be relaxing, possibly even sleeping, in the audience.  When I had finished,  Wing very smoothly jumped up and continued where I left off.  I was impressed: Wing was able to negotiate my contribution to his talk \textit{in his talk} and then incorporated it in a very natural way.  ``Pretty cool!'', I thought.

Second, some remarks.  Most of my contacts with Wing have been  short and spaced-out over many years. As a result, I feel that I don't know him extremely well, but I do know that he is a good guy. I know that he is super-interested in science (especially planetary science), that he works hard, that he has published papers for almost five decades, that he is a little bit mysterious,  and that he has an omnipresent sense of humor.  He is also not a big self-promoter which, in astronomy, is an increasingly unusual attribute.  In fact, those are all good characteristics that are rarely found in combination in any one scientist and  I respect him accordingly.  I also know that Wing spent several years as a university dean and, even worse, a university vice-president.  Being a dean or a vice-president is about as good for the  brain as a kick in the head: very, very few such people ever survive to get back into  science, even if they want to.   Wing seems to be the rare exception, for which he also earns my deep respect.

\section{Summary}
\begin{itemize}
\item Discoveries in the Kuiper belt have prompted a stream of increasingly elaborate, but largely untestable, numerical models of solar system evolution.  While it is clear that we have a much improved appreciation for the complexity of the solar system, there is also an undeniable feeling that the models have  passed the point of diminishing scientific returns. Something must change if we are to make real progress.  
\item Kuiper belt populations, both dynamically hot and cold, include large fractions of objects with ultrared (B-R $>$ 1.6, $S' >$ 25\%/1000\AA) colors, tentatively interpreted as a marker for cosmic ray-processed complex organics.  About 2/3rds of cold-classical KBOs and 1/3rd of the hot populations are ultrared.
\item High perihelion ($q \gtrsim$ 8 to 10 AU) Centaurs closely resemble the Kuiper belt hot population by including a $\sim$1/3rd ultrared matter fraction in their optical color distribution.  
\item Smaller perihelion Centaurs are depleted in ultrared matter, probably as a result of a temperature-dependent evolutionary effect (most likely ballistic resurfacing, given the coincidence between the critical distance for the onset of Centaur activity and the disappearance of the ultrared matter).  
\item Long-period (Oort cloud) and short-period (Kuiper belt) comets are statistically identical in their optical color distributions, with no dependency on perihelion distance or other orbital elements.  All lack ultrared matter.
\item The Jovian Trojan optical colors have an ultrared fraction near 0, inconsistent with a Kuiper belt source.  However, while they are apparently not now active,  ancient Centaur-like activity driven by their first approach to the Sun with $q \lesssim $ 10 AU could have lead to the loss of ultrared mantle material.  
\item The same explanation cannot apply to the Neptunian Trojans. They have the same optical color distribution as found in the Jovian Trojans but they are far too distant and cold for any thermal process to reset the surface colors.  This Trojan color conundrum presently has no obvious resolution.
\end{itemize}


\acknowledgements I thank Ariel Graykowski, Man-To Hui, Pedro Lacerda, Jing Li, and Dave Milewski  for reading the manuscript.



\end{document}